\documentclass[useAMS,usenatbib]{mn2e}
\usepackage{myaasmacros,epsfig}

\def\ltsima{$\; \buildrel < \over \sim \;$}
\def\simlt{\lower.5ex\hbox{\ltsima}}   
\def\gtsima{$\; \buildrel > \over \sim \;$}
\def\simgt{\lower.5ex\hbox{\gtsima}}
\newcommand\bcite[1]{\citeauthor{#1} \citeyear{#1}}

\title[The tidal stripping of satellites]
{The tidal stripping of satellites}

\author[Read et. al.]{J. I. Read $^1$\thanks{Email: jir22@ast.cam.ac.uk} 
M. I. Wilkinson $^1$ N. W. Evans $^1$ G. Gilmore $^1$ \& Jan T. Kleyna $^2$ 
\\ $^1$Institute of Astronomy, Cambridge University, Madingley Road, 
Cambridge, CB3 OHA, England
\\ $^2$Institute for Astronomy, University of Hawaii, 2680 Woodlawn Drive, Honolulu, HI 96822}

\begin{document}

\maketitle

\begin{abstract}
\noindent
We present an improved analytic calculation for the tidal radius of
satellites and test our results against N-body simulations.

The tidal radius in general depends upon four factors: 
the potential of the host galaxy, the potential of the satellite, the
orbit of the satellite and {\it the orbit of the star within the
  satellite}. We demonstrate that this last point is 
critical and suggest using {\it three tidal radii} to cover the
range of orbits of stars within the satellite. In this way we
show explicitly that prograde star orbits will be more easily
stripped than radial orbits; while radial orbits are more easily
stripped than retrograde ones. This result has previously been
established by several authors numerically, but can now be understood
analytically. For point mass, power-law (which includes the
isothermal sphere), and a restricted class of split power law
potentials our solution is fully analytic. For more general
potentials, we provide an equation which may be rapidly solved
numerically.

Over short times ($\simlt 1-2$ \,Gyrs $\sim 1$ satellite orbit), we
find excellent agreement between our analytic and numerical models. Over longer
times, star orbits within the satellite are transformed by the tidal
field of the host galaxy. In a Hubble time, this causes a convergence
of the three limiting tidal radii towards the prograde stripping
radius. Beyond the prograde stripping radius, the
velocity dispersion will be tangentially anisotropic.

\end{abstract}

\begin{keywords}{galaxies: dwarf, galaxies: star clusters, galaxies:
    kinematics and dynamics}

\end{keywords}

\section{Introduction}\label{sec:introduction}

The tidal radius is the radius at which a star within a
satellite becomes unbound or {\it stripped} and becomes bound
instead to the host galaxy about which the satellite orbits. The
problem of calculating the tidal radius of satellites has a long
history which dates back to \citet{1957ApJ...125..451V} who analysed
the problem in the context of globular clusters around the Milky Way. 

It is a problem with a very wide scope of application, from
understanding the sharp edge observed in some nearby globular clusters
\citep{1962AJ.....67..471K}, to semi-analytic modelling of galaxy
formation, where interactions and mergers play a vital role in our
current hierarchical formation paradigm
\citep{2003MNRAS.341..434T}. It is a problem which has also
experienced a renaissance with the advent of so-called `near-field
cosmology'. Our current cosmological models are beginning to give precise
predictions for the distribution, abundance and morphology of
satellite galaxies \citep{1999ApJ...524L..19M}, while data from the
nearest satellites within the Local Group can provide excellent
constraints on these theories
\citep{1998ARA&A..36..435M}. Understanding in detail the effects of a 
tidal field on the stars and dark matter particles within a satellite
are central to these modern branches of astrophysics. 

In this paper, we revisit this old problem and present an improved
analytic calculation for the tidal radius of satellites. It has been
understood, at least numerically, for some time that the tidal radius
in general depends upon four factors: the potential of the host
galaxy, the potential of the satellite, the orbit of the satellite
and, {\it the orbit of the star within the satellite}. To our
knowledge, however, only the first three factors have been addressed
in analytic calculations previously (see
e.g. \bcite{1957ApJ...125..451V} and \bcite{1962AJ.....67..471K}). Here, we
present an analytic calculation which includes the
effect of the orbit of the star too. 

That the orbit of the star is important has been noted several times
in the past. \citet{1972ApJ...178..623T} found numerically that
prograde orbits are more easily stripped than
retrograde orbits, while \citet{1975AJ.....80..290K} showed that both
radial and prograde orbits are more easily stripped than
retrograde orbits. More recently, \citet{2004ApJ...601...37K} and
\citet{2004ApJ...609..482K} have found that the satellites in their
simulations show tangential velocity anisotropy near the tidal radius and
that radially biased velocity distributions are more rapidly
stripped. Similar results have been observed in globular cluster and
star cluster simulations (\bcite{1985IAUS..113..343S},
\bcite{1997MNRAS.292..331T}, 
\bcite{1997MNRAS.286..709G} and \bcite{2003MNRAS.340..227B}); and in
numerical studies of the restricted three-body problem
\citep{1970A&A.....9...24H}. All of
these numerical results point to the fact that the 
orbits of stars play an important role in determining the tidal radius
for the satellite. In this paper, we shed some analytic insight
on these numerical observations.

This paper is organised as follows: in section \ref{sec:theory} we
present our calculation for the tidal radius of
spherical systems. We include the effect of the orbit of the star
within the satellite by considering three limiting cases of interest:
prograde and retrograde circular star orbits, and pure radial star
orbits. In section \ref{sec:numerical}, we then compare our analytic
formulae with detailed N-body simulations of tidal stripping and show
that they provide an excellent fit to the simulation data. Finally, in
section \ref{sec:conclusions}, we present our conclusions.

\section{An analytic calculation of the tidal radius}\label{sec:theory}

\begin{figure}
\begin{center}
\epsfig{file=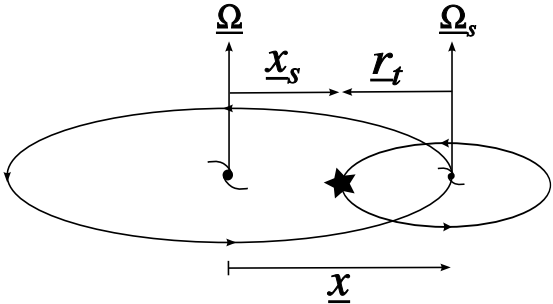,angle=-90,width=8cm}
\caption[]{A schematic diagram of the analytic set-up. The satellite
  and the star within the satellite both orbit in the x-y plane with
  angular velocities $\underline{\Omega}$ and $\underline{\Omega}_s$
  respectively. We consider the problem from a
  frame centred on the host galaxy rotating with angular velocity
  $\underline{\Omega}(t)$. In this frame the centre of mass of the
  satellite is stationary. In this diagram, the star (marked by the black star)
  is on a prograde orbit; see text for further details.}
\label{fig:diagram}
\end{center}
\end{figure}

A schematic diagram of the analytic set-up is shown in Figure
\ref{fig:diagram}. We analyse the problem from a frame 
rotating with angular 
velocity, $\underline{\Omega}(t)$, centred on the host galaxy. We can
choose $\underline{\Omega}(t)$ such that the centre of mass of 
the satellite will always be stationary in this frame. Using
Lagrangian methods as in 
\citet{1987gady.book.....B}, the equation of motion for the centre of
mass of the satellite in this frame is given by:

\begin{equation}
\underline{\ddot{x}} +
\underline{\dot{\Omega}}\wedge\underline{x}+
2\underline{\Omega}\wedge\underline{\dot{x}}+
\underline{\Omega}\wedge(\underline{\Omega}\wedge\underline{x})+
\underline{\nabla}\Phi_g(\underline{x}) = 0
\label{eqn:satmasscent}
\end{equation}

\noindent
where $\underline{x}$ is the vector distance joining the centre of mass of
the satellite to that of the host galaxy, and
$\underline{\nabla}\Phi_g(\underline{x})$ is the force per unit mass due
to the host galaxy  potential, $\Phi_g(\underline{x})$, on the centre of
mass of the satellite.

In the same frame as above, the equation of motion for a star at an
instantaneous distance, $\underline{x}_s$, from the host galaxy
is given by:
  
\begin{equation}
\underline{\ddot{x}}_s +
\underline{\dot{\Omega}}\wedge\underline{x}_s+
2\underline{\Omega}\wedge\underline{\dot{x}}_s+
\underline{\Omega}\wedge(\underline{\Omega}\wedge\underline{x}_s)+
\underline{\nabla}\Phi_g(\underline{x}_s) +
\underline{\nabla}\Phi_s(\underline{r}_t) = 0
\label{eqn:starmasscent}
\end{equation}

\noindent
where $\underline{\nabla}\Phi_s(\underline{r}_t)$ is the force on
the star due to the satellite and $\underline{r}_t =
\underline{x}_s-\underline{x}$ is the instantaneous tidal radius of
the star. Note that from here on we will assume in this calculation
that the star is at its instantaneous tidal radius, $\underline{r} =
\underline{r}_t$.

For the special case of stars on pure circular or pure radial orbits
(two limiting extremes of interest), we may write:

\begin{equation}
\underline{\dot{x}}_s = \underline{\dot{x}} +
\underline{\Omega}_s\wedge\underline{r}_t
\label{eqn:xsdot}
\end{equation}

\noindent
where $\underline{\Omega}_s$ is the angular velocity of the star about
the centre of mass of the satellite. Equation (\ref{eqn:xsdot}) is
valid for circular orbits 
because in this case the space velocity of the star is due only to the
velocity of the satellite galaxy plus that due to its rotation,
$\underline{\Omega}_s$. It is also valid for
pure radial orbits because in this case, not only does $\underline{\Omega}_s
\rightarrow 0$, but crucially, at the apocentre of the star's orbit,
the radial component of the star's space velocity about the satellite
must also be zero. Since all stars on radial orbits with apocentres
larger than the instantaneous tidal radius will be stripped, we may
take the limiting case where radial star orbits at the tidal radius are at
apocentre.

Equating equations (\ref{eqn:satmasscent}) and (\ref{eqn:starmasscent}),
substituting for equation (\ref{eqn:xsdot}), and using the fact that at
the tidal radius,
$\underline{\ddot{r}}_t=\underline{\ddot{x}}_s-\underline{\ddot{x}}=0$,
we obtain the following equation for $r_t$: 

\begin{equation}
\underline{F}_f+\underline{F}=0
\label{eqn:ontheway}
\end{equation}
where
\begin{equation}
\underline{F}=\underline{\nabla}\Phi_g(\underline{x})-\underline{\nabla}\Phi_g(\underline{x}_s)-\underline{\nabla}\Phi_s(\underline{r}_t)
\label{eqn:forceterms}
\end{equation}
\begin{equation}
\underline{F}_f=-\underline{\dot{\Omega}}\wedge\underline{r}_t-\underline{\Omega}\wedge(\underline{\Omega}\wedge\underline{r}_t)-
2\underline{\Omega}\wedge(\underline{\Omega}_s\wedge\underline{r}_t)
\label{eqn:fictforceterms}
\end{equation}

\noindent
Notice that the Coriolis terms,
$2\underline{\Omega}\wedge\underline{\dot{x}}$, have cancelled. For
the case of stars on pure radial orbits where $\underline{\Omega}_s
\rightarrow 0$, all Coriolis terms disappear which is why in the
original calculation by \citet{1962AJ.....67..471K}, such terms can be
left out.

Equations (\ref{eqn:forceterms}) and (\ref{eqn:fictforceterms}) are still
quite unwieldy and give a continuum of tidal radii for stars on
different orbits and different alignments to the orbital plane of the
satellite. We may greatly simplify things, however, by searching for
the smallest tidal radii. This occurs for the co-planar
orientation shown in Figure \ref{fig:diagram}. By placing the star in
the same orbital plane as the satellite, the star's rotational velocity
then maximally adds to (prograde orbits) or subtracts from (retrograde
orbits) the space velocity of the satellite about the host
galaxy. Furthermore, we are interested in the phase of the star's orbit
which similarly minimises the tidal radius. This is when the star
is aligned along the line of centres of the host galaxy and satellite
as shown in Figure \ref{fig:diagram}. At this point the star is both
at its closest approach to the host galaxy and at the point
where its rotational velocity is of maximal effect.

Using the geometry of Figure \ref{fig:diagram} and taking just the
force component along the line of centres of the host galaxy and
satellite (along $\underline{\hat{x}}$), we obtain:

\begin{equation}
\underline{F}_f\cdot\underline{\hat{x}}+\underline{F}\cdot\underline{\hat{x}}=0
\label{eqn:finaleq}
\end{equation}
where
\begin{equation}
\underline{F}\cdot\underline{\hat{x}}=\left.\frac{d\Phi_g}{dx}\right|_x
-\left.\frac{d\Phi_g}{dx}\right|_{|x-r_t|}+\left.\frac{d\Phi_s}{dr}\right|_{r_t}
\label{eqn:forcetermssc}
\end{equation}
\begin{equation}
\underline{F}_f\cdot\underline{\hat{x}} = -\Omega^2 r_t - 2 \alpha\Omega
\Omega_s r_t
\label{eqn:threetidal}
\end{equation}
and
\begin{equation}
\alpha = \left\{\begin{array}{ll}
      1 & \mbox{prograde} \\
      0 & \mbox{radial} \\
     -1 & \mbox{retrograde}
     \end{array}
     \right.
\label{eqn:alpha}
\end{equation}
\noindent
Notice that the terms involving $\underline{\dot{\Omega}}$ vanish in
this geometry and that we have introduced the notation:  $\Omega =
|\underline{\Omega}|$; and similarly for the other vectors.

For the extremal cases of pure circular and pure radial orbits,
we now have three tidal radii for prograde, radial and
retrograde orbits. It is instructive to rearrange equation
(\ref{eqn:finaleq}) to give:

\begin{equation}
\frac{GM_g(x)}{x^2} - \frac{GM_g(x-r_t)}{(x-r_t)^2} +
\frac{GM_s(r_t)}{r_t^2} - \Omega^2 r_t - 2 \alpha \Omega \Omega_s r_t=0
\label{eqn:finaleqrea}
\end{equation}
\noindent
where $M_s(r)$ and $M_g(x)$ are the satellite and host galaxy mass
distributions respectively. Two important points can be made from
equation (\ref{eqn:finaleqrea}). First, the fictitious forces will be
maximised for retrograde orbits and minimised for prograde orbits. Thus, we
expect that prograde orbits are more easily stripped than
radial orbits; while radial orbits are more easily stripped than
retrograde orbits. This will be tested against numerical experiments
in section \ref{sec:numerical}. Secondly, if there were only one tidal
radius (as is the case for $\Omega_s=0$), then there will always be
point mass solutions for $M_g(x)$ and $M_s(r_t)$ which recover this
tidal radius. This is the familiar result that a single tidal radius
contains information only about the total enclosed mass of the
satellite and host galaxy. This is no longer true once the star orbits are
taken into account. In this case $\Omega_s(r_t)^2 = GM_s(r_t)/r_t^3
\neq 0$ and the three limiting tidal radii will depend on the {\it mass
distribution} of the satellite as well as the total enclosed mass. This
point is a direct and important consequence of including the effect of
star orbits in the calculation of the tidal radius.

For spherical symmetry, the energy equation for the satellite may be
rearranged to give:

\begin{eqnarray}
J^2 & = & 2\frac{(\Phi_g(x_p) - \Phi_g(x_a))x_a^2x_p^2}{x_p^2-x_a^2}
\label{eqn:solveJ} \\
E & = & \frac{J^2}{2x_a^2}+\Phi_g(x_a)
\label{eqn:solveE}
\end{eqnarray}
\noindent
where $J$ is the specific angular momentum of the satellite, $E$ is
the specific energy and $x_p$ and $x_a$ are the satellite pericentre
and apocentre respectively. The satellite orbit is then fully
specified by $x_p < x < x_a$ and we may readily obtain $\Omega(t) =
J/x(t)^2$ from equation (\ref{eqn:solveJ}).

Since $\Omega = \Omega(t)$ and $x = x(t)$, for all but circular
satellite orbits, the three limiting tidal radii solutions to equation
(\ref{eqn:finaleqrea}) are a {\it function of time}: $r_t = r_t(t)$.
Whether a time independent value for the tidal radius is appropriate in such
situations is discussed further in section \ref{sec:genorbits}.

\subsection{Point mass potentials}

Given $\Phi_g(x)$, $\Phi_s(r_t)$, $x(t)$, $x_p$ and $x_a$, we now have enough
information to solve equation (\ref{eqn:finaleqrea}) for the three values
of $r_t(t)$. In general, this solution must be found numerically. An
analytic solution may be obtained, however, for the special case of
point mass potentials for the satellite and host galaxy, in which case
the satellite's orbit is Keplerian. In this case, equation
(\ref{eqn:solveJ}) gives us:

\begin{equation}
\Omega^2 = \frac{GM_g}{x^4} \Lambda
\label{eqn:omega2}
\end{equation}
\begin{equation}
\Lambda = \frac{2 x_a x_p}{x_a+x_p} = a(1-e^2)
\label{eqn:kappa}
\end{equation}

\noindent
where $G$ is the gravitational constant, $M_g$ is the mass of the
host galaxy and we define here the standard Keplerian semi major axis,
$a$, and eccentricity, $e$, as:

\begin{equation}
e = \frac{x_a-x_p}{x_a+x_p}
\label{eqn:ecc}
\end{equation}
\begin{equation}
a = \frac{x_a+x_p}{2}
\label{eqn:a}
\end{equation}

\noindent
In fact, these definitions involve only $x_a$ and
$x_p$, and so hold good for any orbit in a spherical potential.

Substituting equation (\ref{eqn:omega2}) into equation
(\ref{eqn:finaleq}), using point mass potentials, and solving, we obtain:

\begin{equation}
r_t \simeq \left(x^4 \frac{M_s}{M_g} \Lambda\right)^{1/3}\left(\frac{\sqrt{\alpha^2+1+2x/\Lambda}-\alpha}{\Lambda+2x}\right)^{2/3}
\label{eqn:rtkepler}
\end{equation}

\noindent
where $M_s$ and $M_g$ are the mass of the satellite and host galaxy
respectively and $\alpha$ is as in equation \ref{eqn:alpha}.

Equation (\ref{eqn:rtkepler}) gives $r_t(x)$ or since $x=x(t)$,
equivalently, $r_t(t)$. As originally suggested by
\citet{1957ApJ...125..451V} and \citet{1962AJ.....67..471K}, we may
take the limiting case which gives the smallest tidal radius -
namely that at pericentre; $x=x_p$. In this case, equation (\ref{eqn:rtkepler})
reduces to:

\begin{equation}
r_t \simeq x_p \left(\frac{M_s}{M_g}\right)^{1/3}\left(\frac{1}{1+e}\right)^{1/3}\left(\frac{\sqrt{\alpha^2+1+\frac{2}{1+e}}-\alpha}{1+\frac{2}{1+e}}\right)^{2/3}
\label{eqn:rtperi}
\end{equation}
\noindent
where $e$ is the eccentricity of the satellite orbit.

It is reassuring to note that, for the case of pure radial
star orbits, $\alpha=0$ and equation (\ref{eqn:rtperi}) reduces to the
familiar \citet{1962AJ.....67..471K} tidal radius:

\begin{equation}
r_t \simeq x_p\left[\frac{M_s}{M_g(3+e)}\right]^{1/3}
\label{eqn:king}
\end{equation}

The above analytic forms for point masses are useful in explicitly
showing that $r_t$ has three limiting values at each position,
$x_p < x < x_a$, of the satellite. From equation (\ref{eqn:rtperi}), we
see that the prograde stripping radius ($\alpha=1$) is smaller
than the radial one ($\alpha=0$) which is smaller than the retrograde
one ($\alpha=-1$), as expected. 

\subsection{Power law potentials}

In practice, point mass potentials are a poor approximation to
observed galaxy mass distributions which are much more
extended. Another set of analytic solutions to equation
\ref{eqn:finaleqrea} may be obtained for power-law density profiles of
the form $\rho_{s,g} = A_{s,g}r^{-\gamma_{s,g}}$, where as previously
$s$ and $g$ subscripts denote the satellite and galaxy
respectively\footnote{Such a density profile is only physical for a
  restricted range of $2 \le \gamma_{s,g} \le 3$
  \citep{1987gady.book.....B}.}. Solving equation \ref{eqn:finaleqrea}
for these density profiles gives:

\begin{equation}
r_t \simeq \left[\sqrt{\frac{A_g\Lambda'(3-\gamma_s)}{A_sx^4}}\left[\alpha+\sqrt{\alpha^2-\frac{1-\gamma_g}{3-\gamma_g}\frac{x^{4-\gamma_g}}{\Lambda'}+1}\right]\right]^{-2/\gamma_s}
\label{eqn:powersol}
\end{equation}
\noindent
where
\begin{equation}
\Lambda' = 2\frac{(x_p^{2-\gamma_g}-x_a^{2-\gamma_g})x_a^2x_p^2}{(3-\gamma_g)(2-\gamma_g)(x_p^2-x_a^2)}
\label{eqn:kappadash}
\end{equation}
\noindent
In the limit $x_p=x_a$ (a circular satellite orbit), $\Lambda' =
\frac{x_p^{4-\gamma_g}}{3-\gamma_g}$ and so is well-behaved.

A particular case of interest is for isothermal sphere density
profiles where $\gamma_{s,g}=2$. This is because these approximate
well the observed mass distributions in galaxies. In this special
case, equation \ref{eqn:powersol} reduces to the following simple
form:

\begin{equation}
r_t \simeq \frac{\sqrt{\Lambda''A_s/A_g}\left[-\alpha+\sqrt{\alpha^2+1+x^2/\Lambda''}\right]}{1+\Lambda''/x^2}
\label{eqn:isosol}
\end{equation}
where
\begin{equation}
\Lambda'' = 2\frac{x_a^2x_p^2}{x_p^2-x_a^2}\ln{\frac{x_p}{x_a}}
\label{eqn:kappaddash}
\end{equation}
\noindent
In the limit $x_p=x_a$, $\Lambda'' = x_p^2$. 

For the isothermal sphere, $A_s/A_g$ is more usually written
$\sigma_s^2/\sigma_g^2$, where $\sigma_{s,g}$ is the velocity
dispersion of the satellite/host galaxy. For satellites on pure
circular orbits ($x_a=x_p$), with stars on pure radial orbits
($\alpha=0$), equation \ref{eqn:isosol} reduces to: $r_t =
\frac{x_p\sigma_s}{\sqrt{2}\sigma_g}$, in agreement with the formula
derived recently by \citet{2005Natur.433..389D}.

Another interesting limit to equation \ref{eqn:powersol} is for
$\gamma_{s,g}=3$. This limit recovers the result for point mass
potentials given in equation \ref{eqn:rtkepler}.

\subsection{Split power law potentials}

Power law potentials provide a useful class of analytic
solutions. However, all valid power law solutions except the point mass have
diverging mass at large radii and a limited range of applicability.
(Recall that $2 \le \gamma_{s,g} \le 3$ in these potentials.) Split
power law density profiles do not suffer from these restrictions
(\bcite{1990ApJ...356..359H}, \bcite{1992MNRAS.254..132S},
\bcite{1993MNRAS.265..250D} and \bcite{1996MNRAS.278..488Z}). A fully
analytic sub-set of general split power law profiles is given by:

\begin{equation}
\rho_\mathrm{s,g} = \frac{M_{s,g}(3-\gamma_{s,g})}{4\pi r_{s,g}^3}\frac{1}{(r/r_{s,g})^{\gamma_{s,g}}(1+r/r_{s,g})^{4-\gamma_{s,g}}}
\label{eqn:rhosp}
\end{equation}
\begin{equation}
\Phi_\mathrm{s,g} = \frac{M_{s,g} G}{r_{s,g}(2-\gamma_{s,g})}\left[(1+r_{s,g}/r)^{\gamma_{s,g}-2}-1\right]
\label{eqn:psisp}
\end{equation}
\noindent
where $M_{s,g}$ and $r_{s,g}$ are the mass and scale length in both cases, and
$\gamma_{s,g}$ is the central log-slope of the split power law density
profile.

Equation \ref{eqn:finaleqrea} may be solved using the above split
power law profiles, but is only fully analytic for $\gamma_s=0$ and
$r_t \ll x+r_g$. The second restriction is nearly always an excellent
approximation since it is likely that both $r_t \ll x$ and $r_t \ll
r_g$. The first restriction works for cored satellite halos, and for
satellite halos with $r_t \gg r_s$ for which only the outer regions
are being sampled. In these limits, equation
\ref{eqn:finaleqrea} gives:

\begin{equation}
r_t \simeq \left(x^4\frac{M_s}{M_g}\Lambda'''\right)^{1/3}\left(\frac{\sqrt{\alpha^2+1-qx^4/\Lambda'''}-\alpha}{\Lambda'''-qx^4}\right)^{2/3}-r_s
\label{eqn:spsol}
\end{equation}
\noindent
where
\begin{equation}
q = \frac{x^{-\gamma_g}}{(x+r_g)^{3-\gamma_g}}\left[1-\gamma_g+\frac{\gamma_g-3}{1+r_g/x}\right]
\end{equation}
\begin{equation}
\Lambda''' = \frac{2x_a^2x_p^2}{r_g(2-\gamma_g)(x_p^2-x_a^2)}\left[(1+r_g/x_p)^{\gamma_g-2}-(1+r_g/x_a)^{\gamma_g-2}\right]
\end{equation}
\noindent
In the limit $x_p=x_a$, $\Lambda''' =
x_p\left(\frac{x_p}{x_p+r_g}\right)^{3-\gamma_g}$. In the limit $r_t
\gg r_s$ and $x \gg r_g$, equation \ref{eqn:spsol} reduces to equation
\ref{eqn:rtkepler} for point mass potentials, as expected.

In the following section,
where we compare the theoretical tidal radii with N-body simulations,
we solve equation (\ref{eqn:finaleqrea}) numerically for the same
potentials as those used in the simulations. Using point mass
potentials instead can lead to errors in the tidal radii as large as a
factor of two; using isothermal potentials would give smaller errors,
since these well-approximate the galactic potentials used in the
simulations.

\begin{figure*}
\begin{center}
\epsfig{file=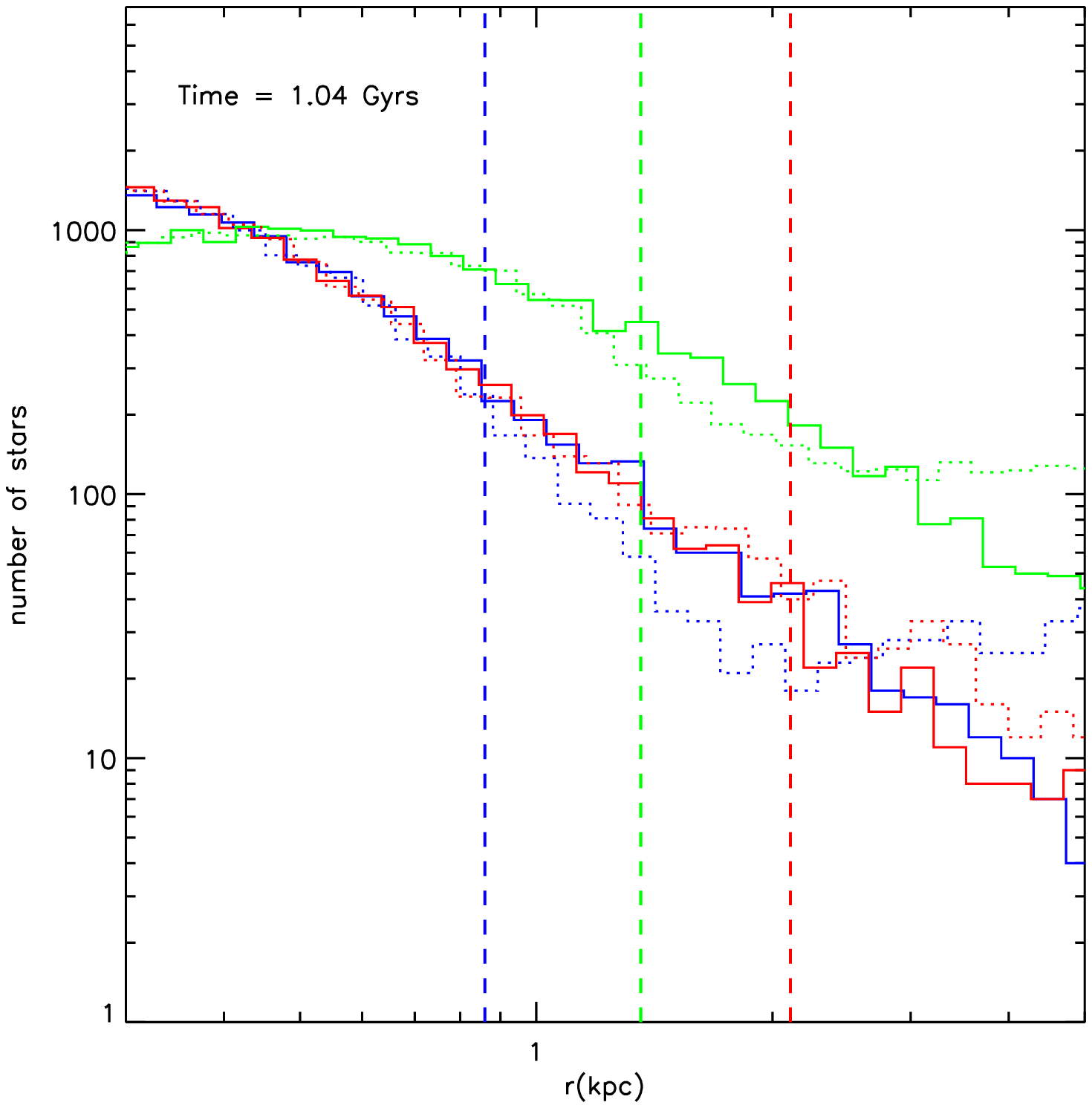,width=8cm}
\epsfig{file=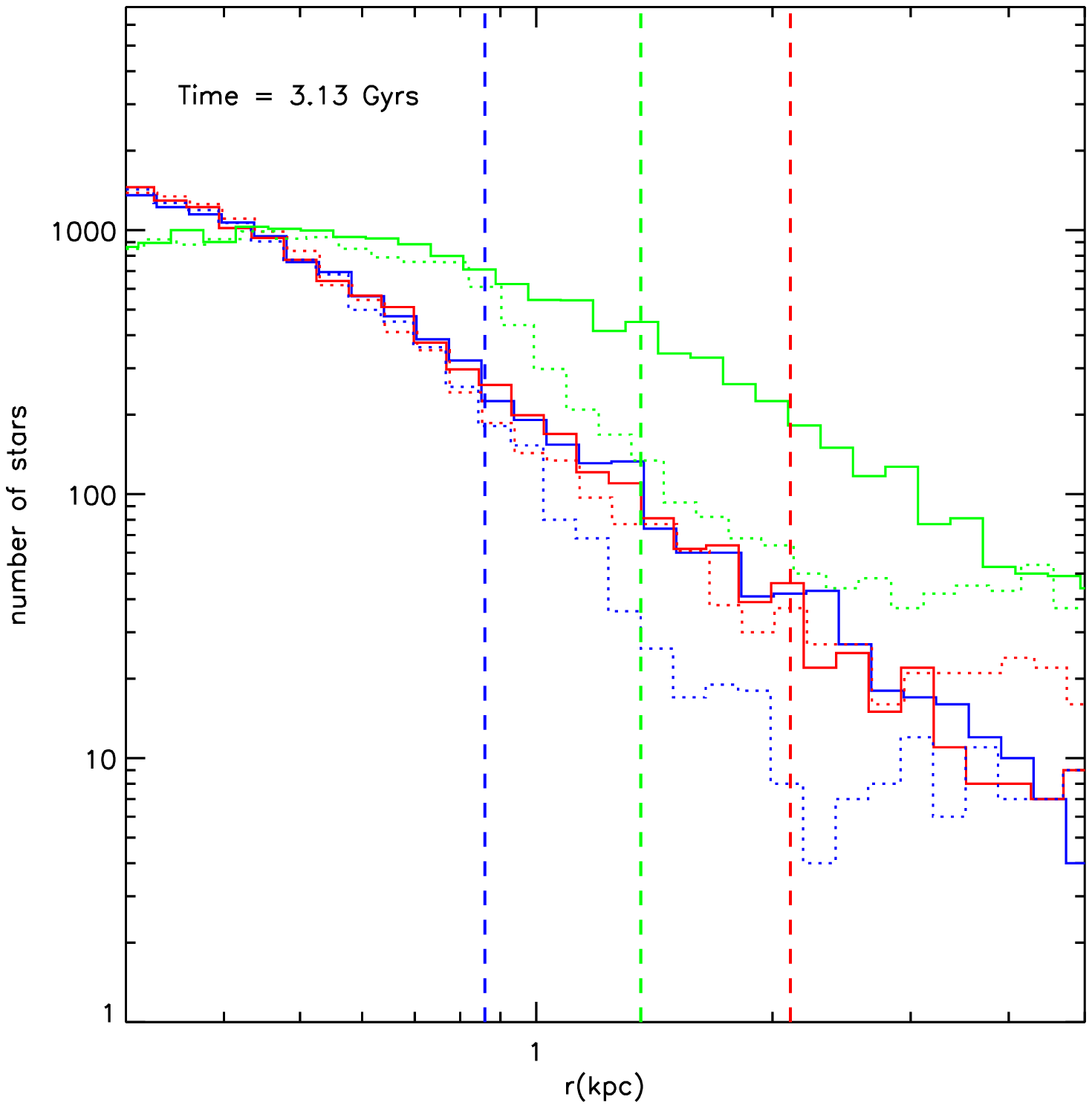,width=8cm}\\
\epsfig{file=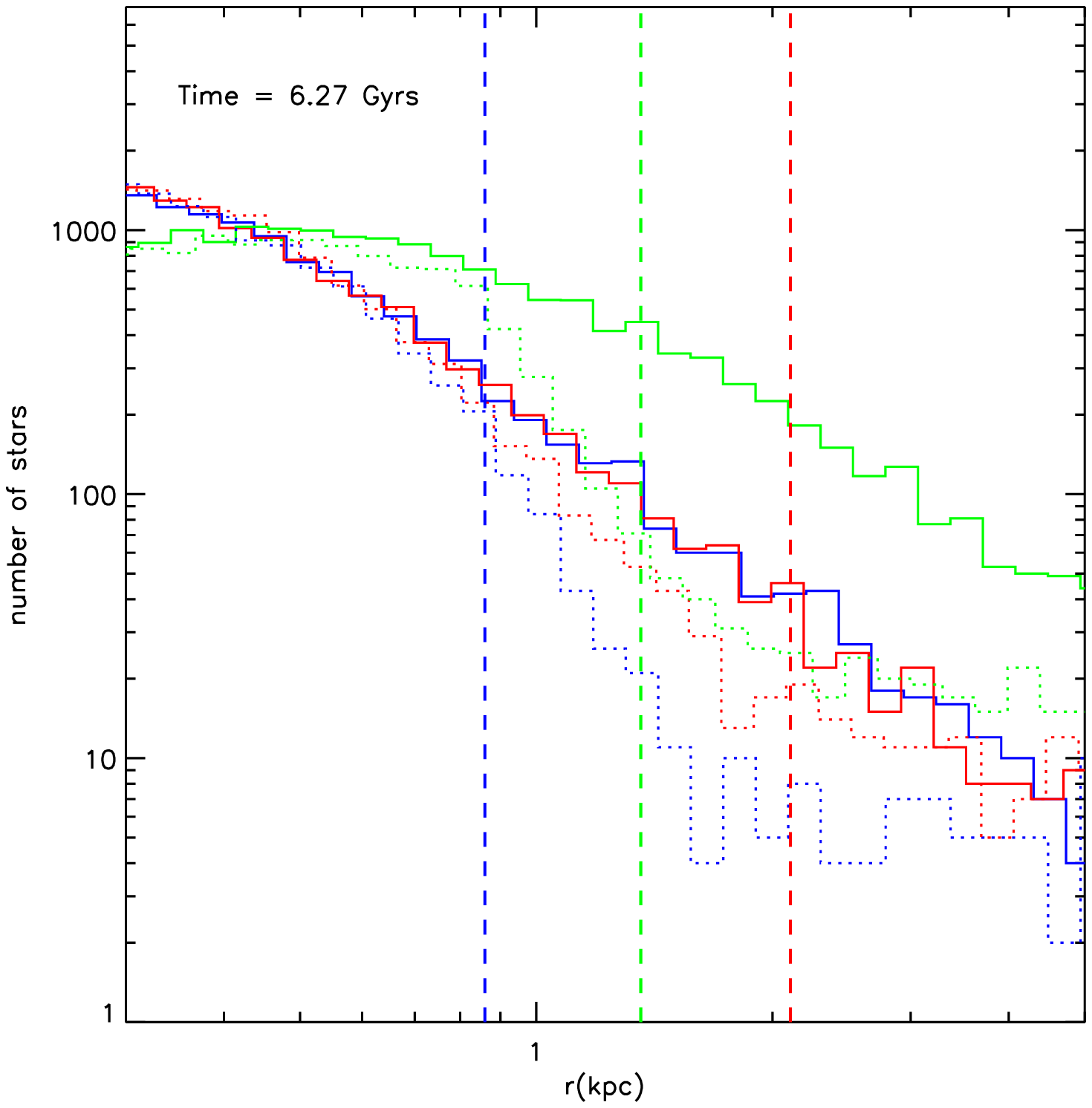,width=8cm}
\epsfig{file=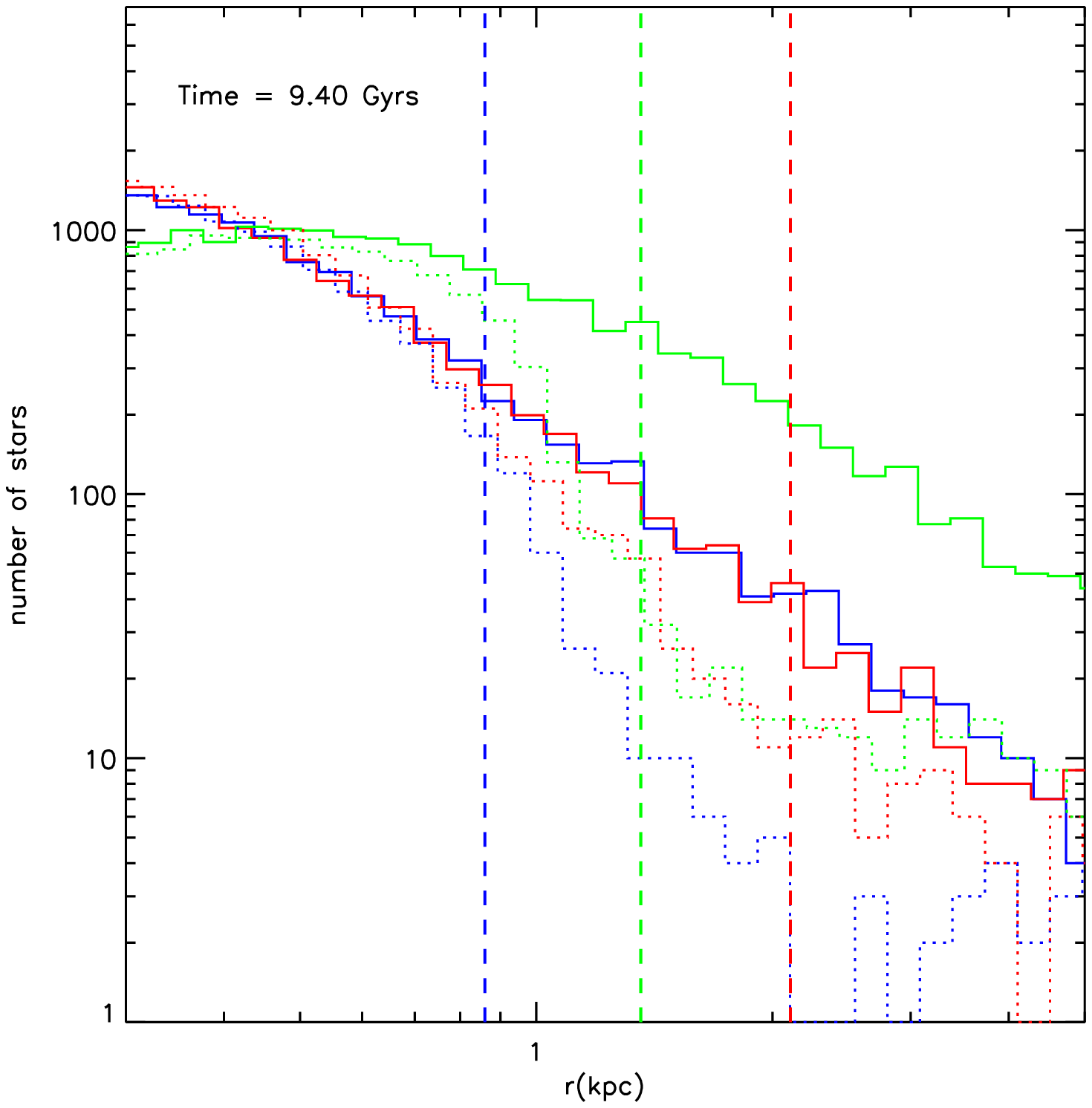,width=8cm}
\caption[]{The tidal stripping of satellite stars as a function of the
stellar orbits. Each panel shows a different time output as the
satellite orbits around the host galaxy on a circular orbit at
80\,kpc. The blue, green and red lines show the
radial distribution of stars which were initially on prograde,
radial and retrograde orbits respectively. The solid lines show the initial
distributions, while the dotted lines show the evolved profiles. The
vertical dashed lines show the three analytic tidal radii
calculated from equation \ref{eqn:finaleq}: prograde (blue), radial
(green) and retrograde (red).}
\label{fig:circorb}
\end{center}
\end{figure*}

\section{Comparison with numerical simulations}\label{sec:numerical}

We set up the initial conditions for the satellite by
drawing the positions from an analytic density profile and the
velocities from a numerically calculated distribution function. This
process is described and tested in detail in a companion paper
\citep{Readinprep1}. We note here that using distribution
functions rather than the more familiar Maxwellian approximation (as
in \bcite{1993ApJS...86..389H}) for calculating particle velocities is
much more accurate for simulations of tidal stripping
\citep{2004ApJ...601...37K}.

We used a Plummer density profile for the satellite which is a
self-gravitating, single component, spherical galaxy comprising only
stars. The Plummer profile is given by \citep{1987gady.book.....B}:

\begin{equation}
\rho_\mathrm{plum} = \frac{3M_s}{4\pi a_s^3}\frac{1}{(1+\frac{r^2}{a_s^2})^{5/2}}
\label{eqn:rhoplum}
\end{equation}
\noindent
where $M_s = 10^7$M$_\odot$ and $a_s=0.23$\,kpc are the mass and scale
length of the satellite respectively. The satellite velocity
distribution was initially isotropic and when evolved in
isolation was found to be extremely stable over a Hubble time. We used
a simulation resolution of $10^5$ particles for the satellite and a
force softening of 10 parsecs; the simulations were found to be very
well resolved - higher resolution test runs with $10^6$ particles and
parsec scale force softening produced converged results.

The initial conditions were evolved using a modified version of the
GADGET N-body code \citep{2001NewA....6...79S} modified to permit a
fixed potential to model the host galaxy. We used a
host galaxy potential chosen to provide a good fit to the Milky Way
\citep{2005ApJ...619..807L}. We used a
Miyamoto-Nagai potential for the Milky Way disc and 
bulge and a logarithmic potential for the Milky Way dark matter
halo. These are given by respectively (see
e.g. \bcite{1987gady.book.....B}):

\begin{equation}
\Phi_\mathrm{mn}(R,z) = \frac{-G M_d}{\sqrt{R^2+(a+\sqrt{z^2+b^2})^2}}
\label{eqn:miyamoto}
\end{equation}
\noindent
where $M = 5 \times 10^{10}$M$_\odot$ is the disc mass, $a=4$\,kpc is
the disc scale length and $b=0.5$\,kpc is the disc scale height, and:

\begin{equation}
\Phi_{\mathrm{log}}(r) =
\frac{1}{2}v_0^2\ln\left(R_c^2+r^2\right)+\mathrm{constant} 
\label{eqn:loghalo}
\end{equation}
\noindent
where $R_c=4.1$\,kpc is the halo scale length 
and $v_0=220$km/s is the asymptotic value of the circular speed of
test particles at large radii in the halo.

We held the satellite properties
and the host galaxy properties fixed and varied only the orbit of the
satellite about the host galaxy. We use a circular orbit at 80\,kpc
(section \ref{sec:circorbits}) and an eccentric orbit with apocentre
85\,kpc and pericentre 23\,kpc (section \ref{sec:genorbits}). These orbits
are typical of Local Group dSph galaxies (see
e.g. \bcite{2002AJ....124.3198P} and \bcite{2005AJ....130...95P}).

Finally, it is not trivial to mass and momentum centre the satellite
when performing analysis of the numerical data. An incorrect mass
centre can lead to spurious density and velocity features
\citep{Readinprep3}. We use the method of shrinking spheres to
find the mass and momentum centre of the satellite
\citep{2003MNRAS.338...14P}.

\subsection{Circular satellite orbits}\label{sec:circorbits}

The first numerical test is a circular orbit of a satellite which experiences
relatively weak tidal effects from the host galaxy. This is
because the analytic formulae assume that the satellite potential
remains constant. If a large fraction of the satellite's mass is
stripped away, this approximation starts to break down. In this
case, the tidal radii should be recalculated at each time step
iteratively as mass is removed, much as is done in semi-analytic
simulations of tidal stripping (see
e.g. \bcite{2003MNRAS.341..434T}).

\begin{figure*}
\begin{center}
\epsfig{file=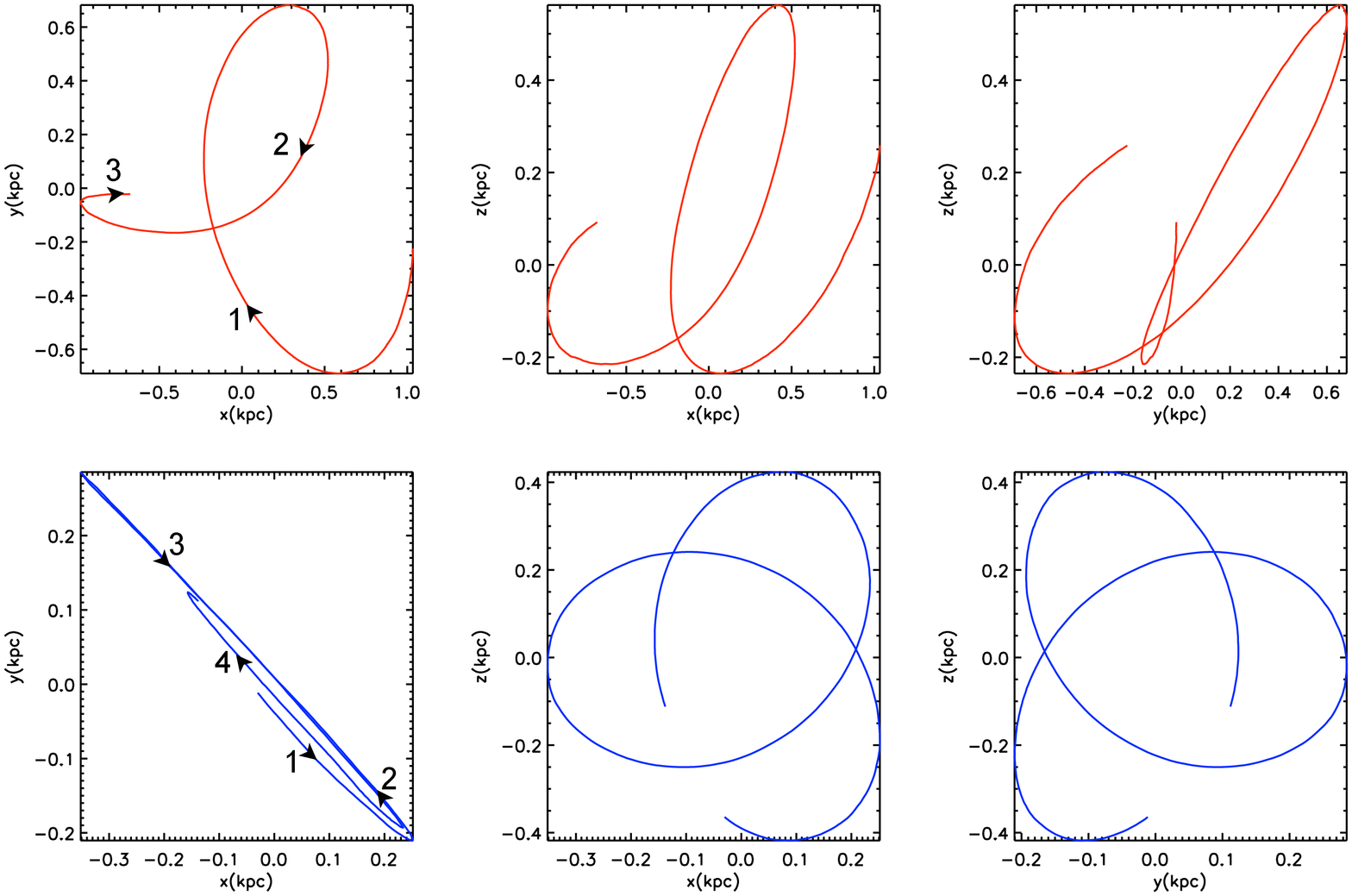,angle=-90,width=16cm}
\caption[]{Space projections for two star orbits which were converted
  to retrograde motion (red line) and to prograde motion (blue line)
  over $\sim 1$ Gyr. Orbits were extracted from the simulation for a
  satellite on a circular orbit about the host galaxy (see section
  \ref{sec:circorbits} and text for more details). The black arrows
  and numbers mark the direction of motion of the star in each case.}
\label{fig:convertorbits}
\end{center}
\end{figure*}

We wish to understand how a satellite star is affected by the host
galaxy's tidal field, as a function of the star's orbit. In order to do
this, we calculate the orbit of each star in the satellite
initially. For a spherical system, the energy
equation for an individual star can be simply rearranged to give
\citep{1987gady.book.....B}: 

\begin{equation}
\dot{r}^2 = 2(E_s-\Phi_s(r)) - J_s^2/r^2
\label{eqn:energyeq}
\end{equation}
where $\dot{r}$ is the radial velocity of the star with respect to the
satellite's centre of mass, $E_s$ is the star
specific energy, $J_s$ is the star specific total angular momentum and
$\Phi_s(r)$ is the potential of the satellite. 

\begin{figure*}
\begin{center}
\epsfig{file=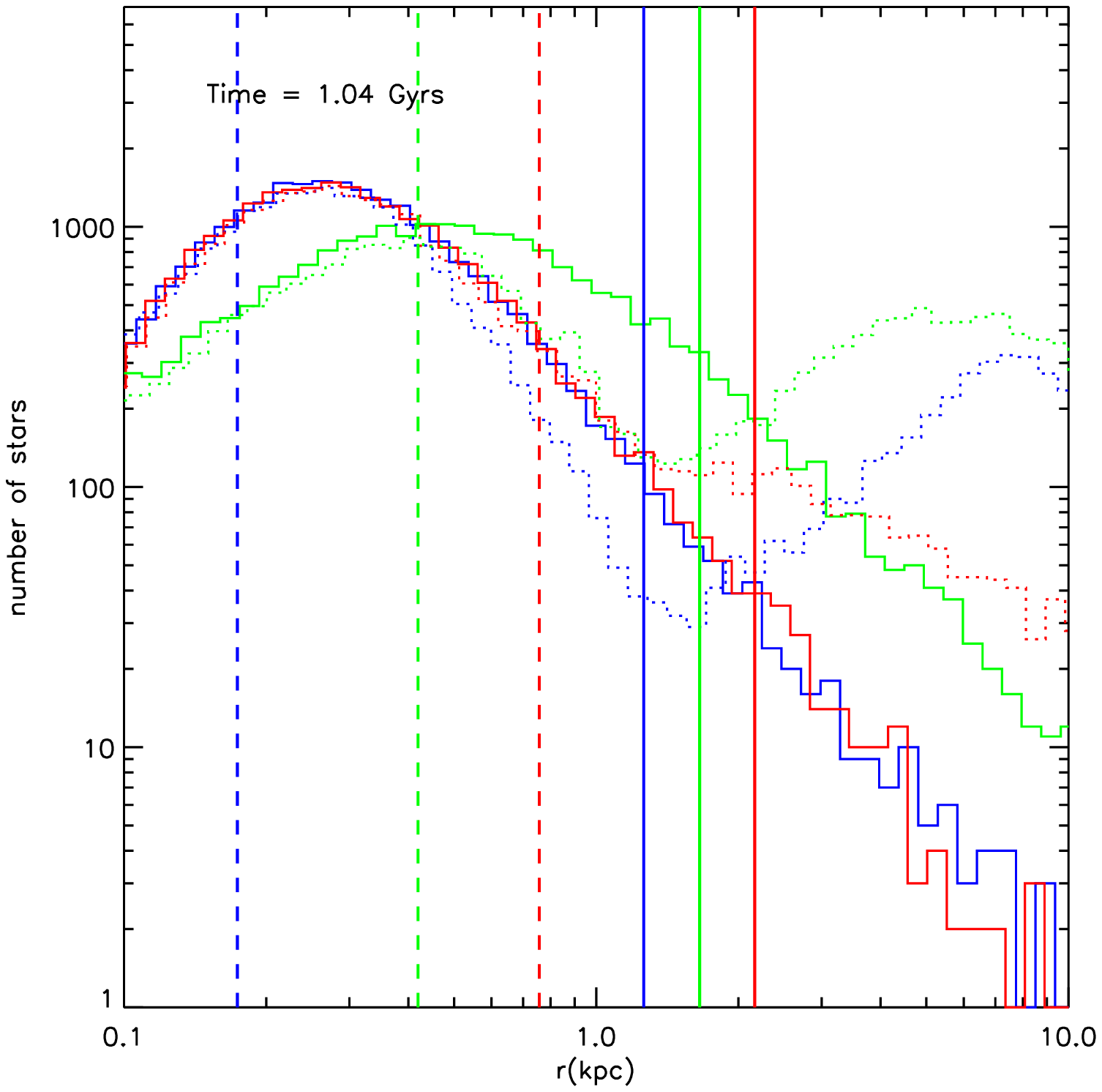,width=8cm}
\epsfig{file=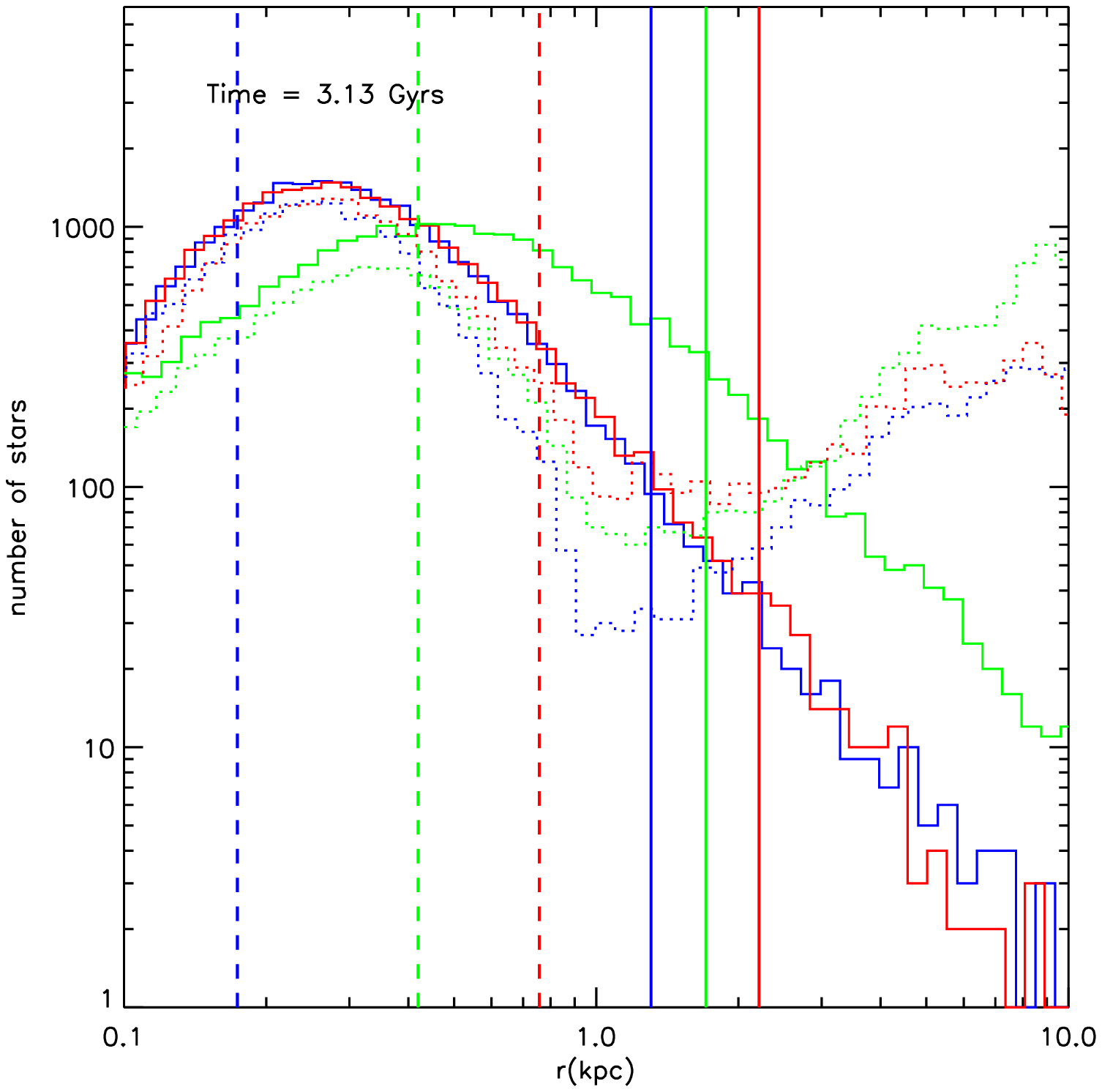,width=8cm}\\
\epsfig{file=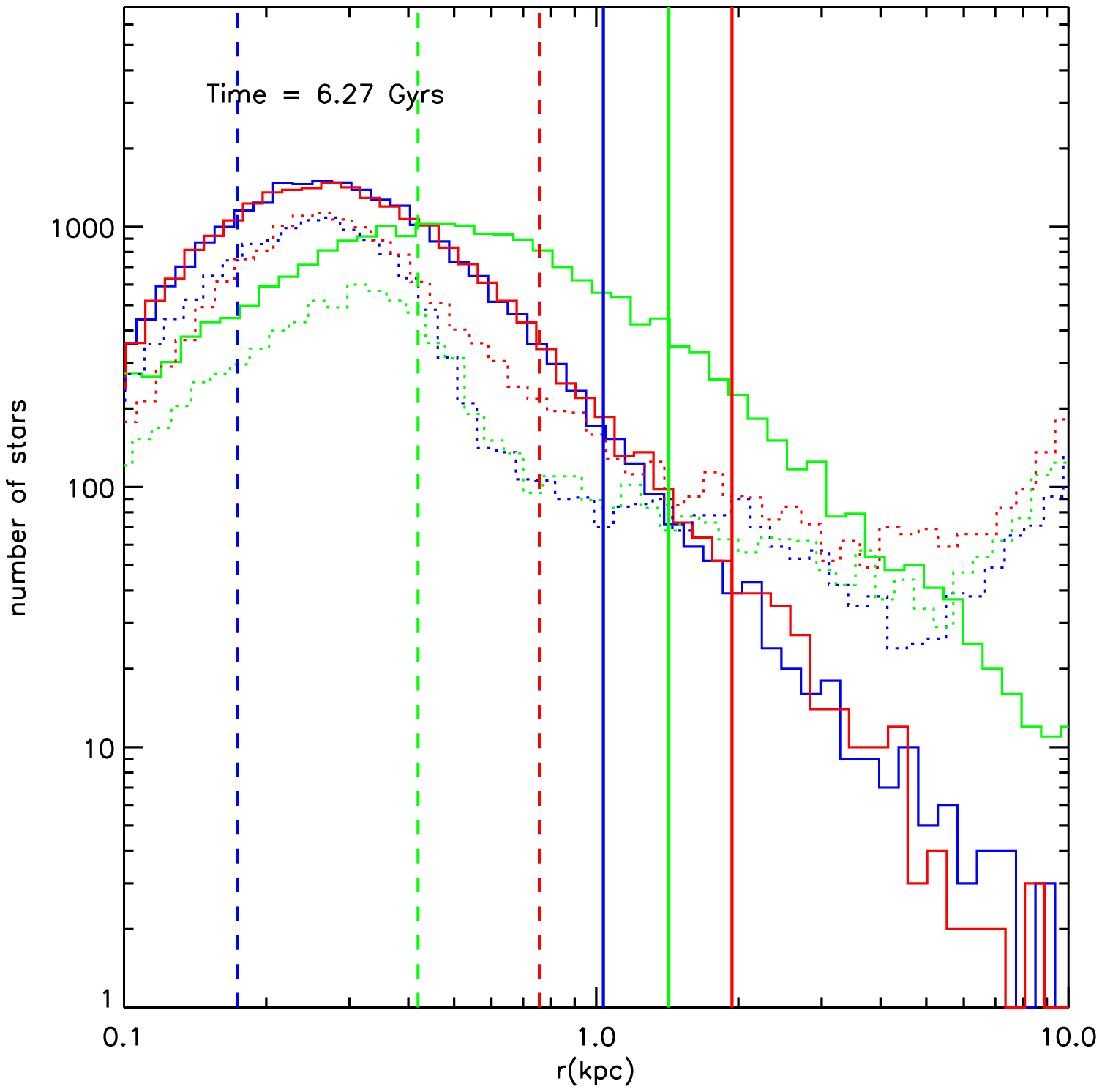,width=8cm}
\epsfig{file=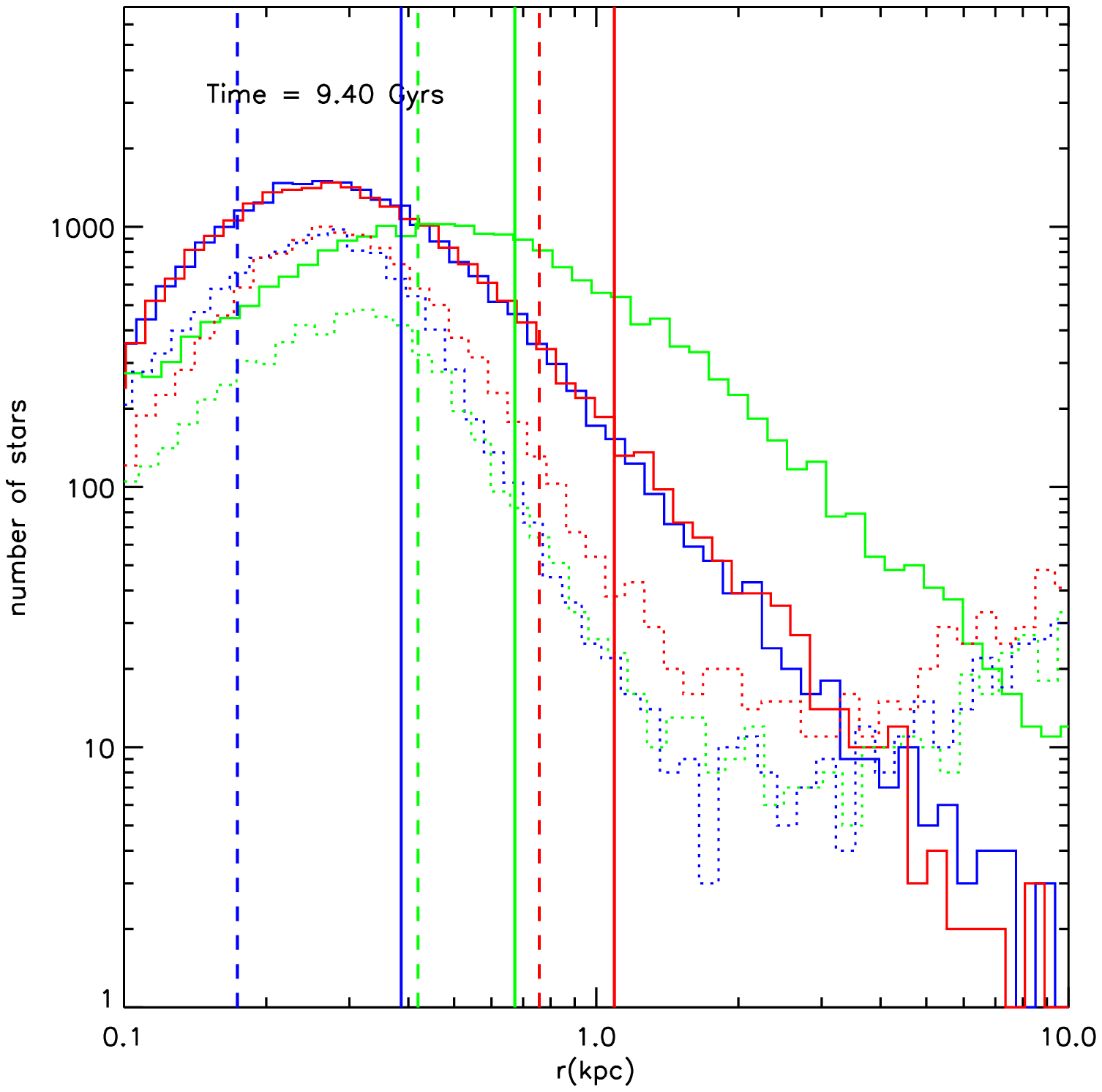,width=8cm}
\caption[]{The tidal stripping of satellite stars as a function of the
stellar orbits. Lines and panels are as in figure \ref{fig:circorb},
but now the satellite is on a highly eccentric orbit with
apocentre 85\,kpc and pericentre 23\,kpc. The vertical solid lines
show the three tidal radii calculated at the current time; while the vertical
dashed lines show the three tidal radii calculated at the pericentre
(distance of closest approach) of the satellite's orbit.}
\label{fig:genorb}
\end{center}
\end{figure*}

At pericentre, $r_p$, and apocentre, $r_a$, $\dot{r}=0$ and so we may
calculate $r_a$ and $r_p$ for each satellite star by finding the
roots of equation (\ref{eqn:energyeq}). We may then obtain the
eccentricity of each star orbit, $e$, as defined in equation
(\ref{eqn:ecc}). In order to do this, we must know $\Phi_s(r)$; and $J_s$
and $E_s$ for each star. The specific kinetic energy and angular momentum for
each star may be simply computed from their phase space
coordinates. For $10^5$ star particles, the potential due to the
self gravitating stellar distribution is prohibitively slow to
compute via direct summation. To avoid this problem, we use a tree
code to compute the potential \citep{2000ApJ...536L..39D}.

Ideally, we would like to compare pure circular and pure radial orbits
with the analytic calculation in section \ref{sec:theory}. However,
models built either with pure circular or pure radial orbits are
unstable (see e.g. \bcite{1987gady.book.....B}). We use instead
an isotropic distribution of velocities which has the advantage that
it is stable, but the disadvantage that very few stars are on either
pure circular or pure radial orbits. To obtain good statistics, we
define an orbit as being `circular' if $e < 0.5$ and `radial' if $e >
0.7$. The `circular' orbits are then divided into prograde
($\underline{J}_s\cdot\underline{J}>0$) and retrograde
($\underline{J}_s\cdot\underline{J}<0$) orbits; where $\underline{J_s}$ and
$\underline{J}$ are the star and satellite specific angular momentum
vectors respectively. The radial orbit boundary ($e > 0.7$) is chosen
such that there are as many stars on `radial' orbits initially as
there are on prograde or retrograde orbits.

Having divided up the initial stellar distribution in this way, we can
track the evolution of prograde, radial and retrograde stars as the
satellite orbits around the host galaxy. Figure \ref{fig:circorb}
shows four snapshots in time over 10\,Gyrs as the satellite orbits
around the host galaxy.

\begin{figure*}
\begin{center}
\epsfig{file=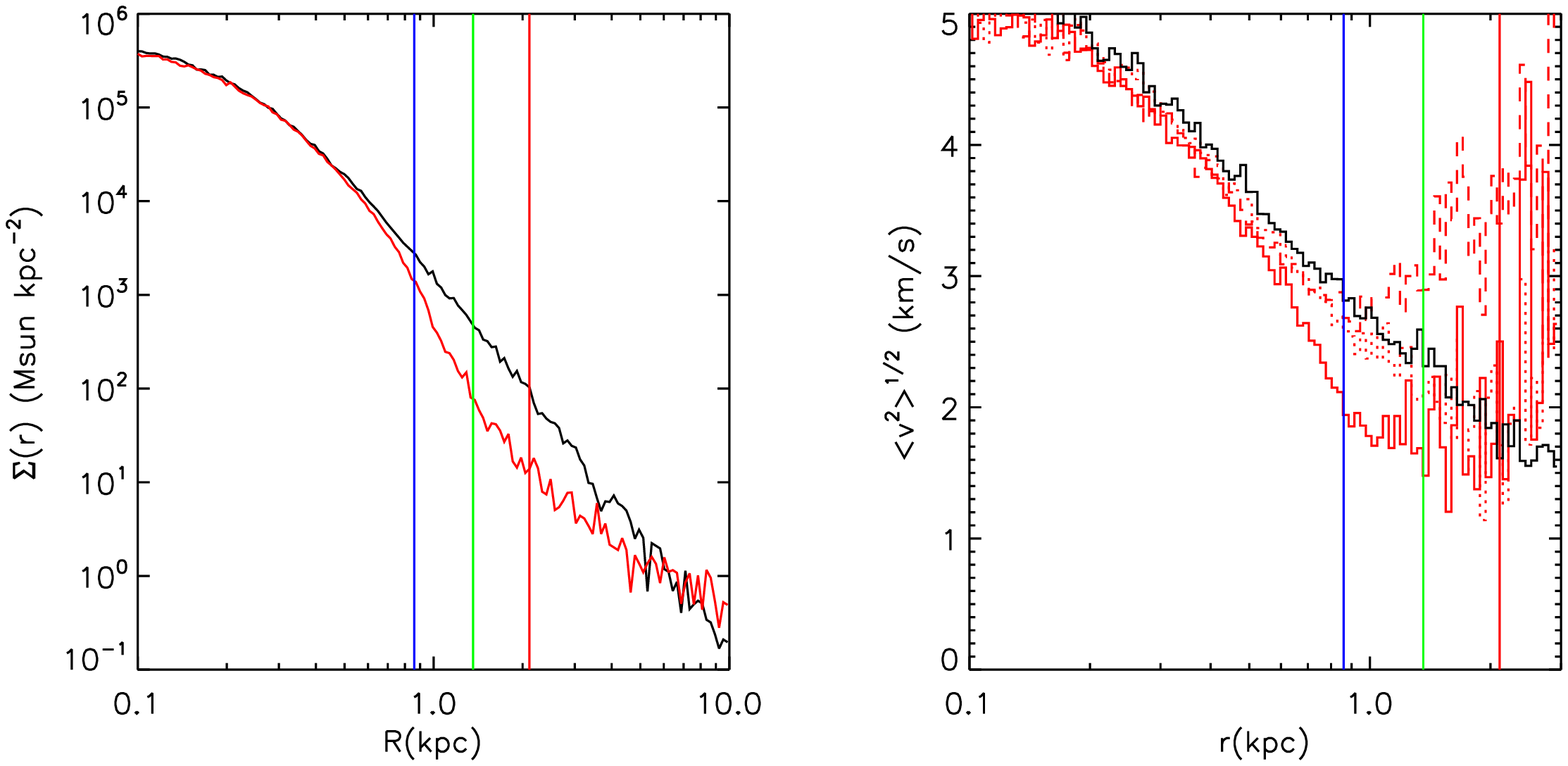,width=16cm}
\epsfig{file=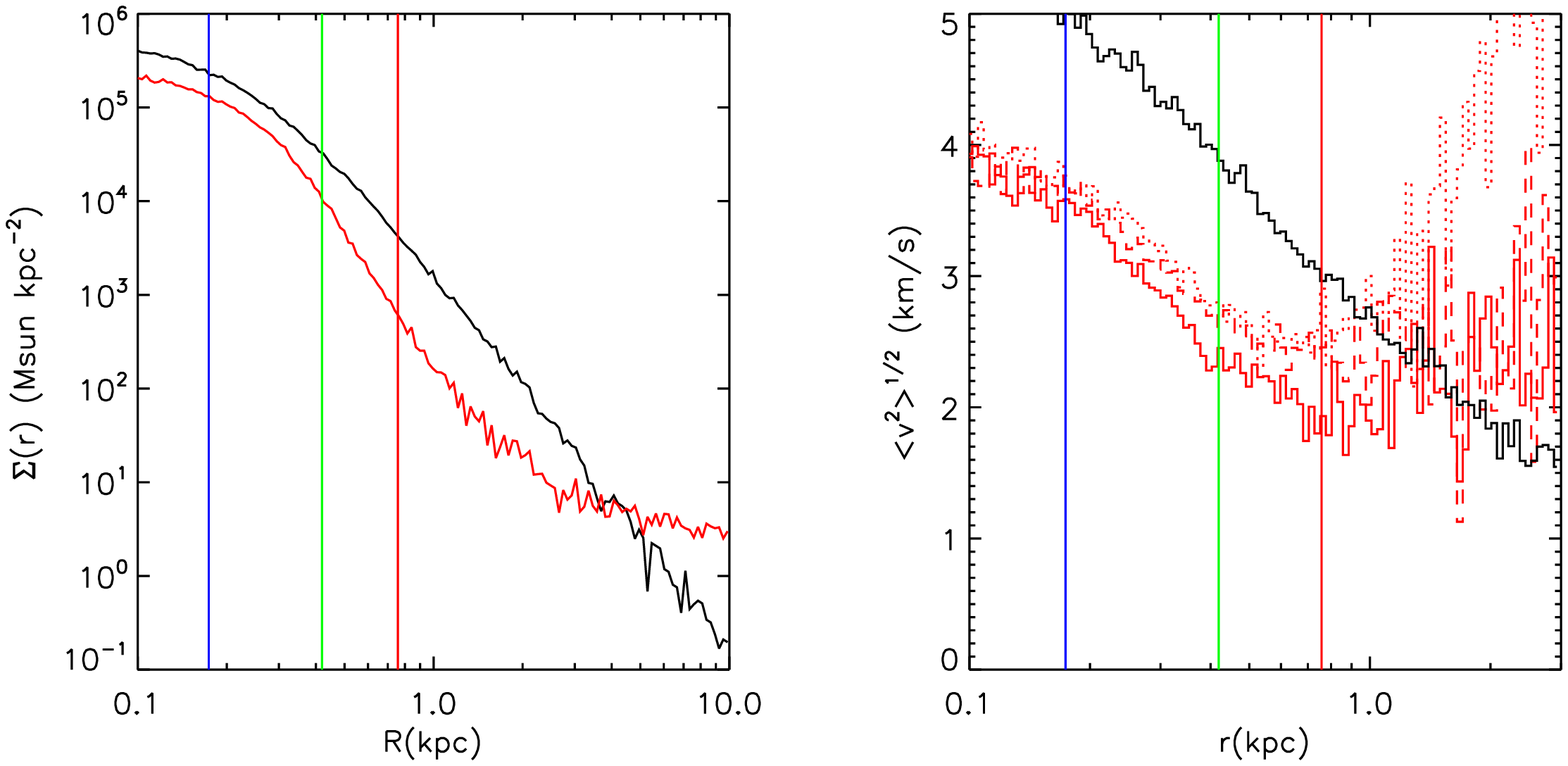,width=16cm}\\
\caption[]{The projected surface brightness distribution (left) and velocity
  dispersion (right) for a satellite on a circular orbit (top panels and see
  section \ref{sec:circorbits}) and an eccentric orbit (bottom panels
  and see section \ref{sec:genorbits}). The black lines show the
  initial profiles while the red lines show the profiles after $\sim
  10$\,Gyrs. The solid vertical red, green and blue lines show the
  three analytic tidal radial from equation \ref{eqn:finaleq}. In the
  plot of the velocity dispersions, the solid red line shows the
  radial velocity dispersion while the dotted and dashed lines show
  the $\theta$ and $\phi$ components.}
\label{fig:realprofiles}
\end{center}
\end{figure*}

From Figure \ref{fig:circorb}, we can see that the analytic model
works very well for the first Gyr. The dotted lines show that
stripping occurs for each of the orbit-types only up to the analytic
tidal radii. Note that we have included both bound and unbound stars
in this plot. This is why the dotted lines turn upwards at large
radii. It is a well-known result that unbound stars can take a
long time to actually drift away. However, that is a different effect
to the one being discussed here. The retrograde stars interior to the
retrograde stripping radius are bound energetically and not just
unbound stars taking a long time to escape. We have explicitly
calculated energies to check this.

Over longer time scales\footnote{The time for
  one circular orbit at 80\,kpc in the potential we use is $\sim
  2$\,Gyrs. Throughout this paper we refer to `short' times as
  being $\simlt$ 2\,Gyrs and `longer times' as being $\simgt 2$\,Gyrs.},
the analytic limits appear to fare less well: all of the orbit types
  begin to become stripped into the prograde limit (blue dashed
  line). The natural question is then, what is causing this effect? 
Two obvious possibilities are that either the potential of the
satellite has changed significantly, in which case the tidal radii
should migrate inwards, or that the orbits of the stars have been altered
by the tidal field of the host galaxy. By plotting the potential of
the satellite as a function of time, it is easy to show that option
one is not correct. As might be expected given the small
number of stars which are actually stripped in this simulation, the
satellite potential remains almost constant over the whole simulation
time of 10\,Gyrs. Rather, the orbits of the stars are
changing in the tidal field of the host galaxy. We discuss this
further in the following section.

\subsection{Star orbits in a tidal field}\label{sec:tidalbraking}

So far we have implicitly assumed that the orbits of stars within the
satellite will be unchanged by the tidal field. This is not the case,
and in fact determining the nature of such orbits within a tidal field
is an interesting problem in its own right which is beyond the scope
of this present work. However, what matters for determining the tidal
radius is only the relative velocity of the star with respect to the
centre of mass of the satellite. In this sense, our three limiting
tidal radii for prograde circular, pure radial and retrograde circular
orbits delimit regions of phase space which, if a star ever enters,
then it will become stripped.

In Figure \ref{fig:convertorbits}, we plot two star orbits taken from the
simulations. If the star orbits were evolved in isolation from a tidal
field, then the star angular momentum would be conserved (since the
satellite potential is spherical). In a tidal field, the star
angular momentum, $J_s$, is not conserved. The z-component, $J_{s,z}$,
can alter, causing a flip from retrograde to prograde motion. Such
stars can thus become stripped even if they start out on orbits which,
if evolved in tidal isolation, would not be. Stars most likely to be
affected in this way are those with low $J_{s,z}$ - those on 
near-polar, or highly eccentric orbits.

In the first orbit (top three space projections in Figure
\ref{fig:convertorbits}), tidal forces
act along the direction of motion of the star. In this case, an
initially prograde orbit loses its tangential velocity
and is converted to a radial orbit. After further deceleration, the
final orbit is only very slightly retrograde and appears nearly radial in the
plot. In the second orbit (bottom three space projections), the star
is on a more polar orbit. Tidal forces acting perpendicular to star's
orbit plane tilt the orbit about the z-axis, causing a flip from
retrograde to prograde motion. Both orbits have low $J_{s,z}$: the
first because it is eccentric; the second because it is polar.

The result of these orbital transformations is that the three tidal
radii of section \ref{sec:theory},
over long times, converge on the prograde stripping radius. This is
because any star beyond the prograde stripping radius which has its
orbit transformed to be prograde is rapidly stripped. Over long
times it becomes increasingly likely that low angular momentum, high
eccentricity, stars will for a short time be pushed onto prograde
orbits, at which point they will be carried away by tides.

\subsection{General satellite orbits}\label{sec:genorbits}

In this section, we consider the more complicated case of a
satellite on a highly eccentric orbit. We use an orbit with apocentre
85\,kpc and pericentre 23\,kpc. Figure \ref{fig:genorb}
shows four snapshots in time over 10\,Gyrs.

As discussed in section \ref{sec:theory}, the three limiting tidal
radii for prograde (blue dashed line), retrograde (red dashed line)
and radial (green dashed line) are functions of
  time. This is why the solid lines are in different places in the
panels of Figure \ref{fig:genorb}. The dashed vertical lines in Figure
\ref{fig:genorb} show the values of the three tidal radii at the
pericentric distance of the satellite; these do not vary with time. An
added complication is that the galactic potential we use is not
spherical. As a result, the orbit of the satellite will precess and
its angular momentum vector, $\underline{J}$, will change
direction. Thus defining `prograde' and `retrograde' stars becomes
formally more difficult as such terms now become a function of time
along the satellite's orbit. With the simulation parameters chosen
here, however, the precession of the satellite's orbit plane is small
over 10\,Gyrs and makes little difference to the initial orbit
classification. Because of this, we use the initial value of
$\underline{J}$ when defining prograde, radial and retrograde stars.

Notice that there are now two points at which the satellite stars show
a tidal `break radius'. One is at the current tidal radius, the other
is close to the pericentric tidal radius. Similar results for the
surface brightness distributions have been observed by other authors (see
e.g. \bcite{2002AJ....124..127J}). Each of these break radii are split into
three further limiting tidal radii as a function of the star orbits. 

Over long timescales ($\simgt 8$\,Gyrs), the break radius which lies at the
current tidal radius of the satellite's orbit disappears. This is
because, after the initial distribution has been stripped of all stars
beyond their tidal radii, further stripping can only occur as stars
heated by the tidal field of the host galaxy migrate outwards. This
effect is much smaller and produces only a tiny perturbation on the
surface brightness profile which, at these late times, now appears
much smoother.

Notice that over a Hubble time, the central density of all 
orbits lowers in its normalisation, even in the very centre of the
satellite. This is not due to tidal 
stripping, but rather to a different physical effect which occurs only
for highly eccentric orbits, or for orbits which cause the satellite
to pass through the plane of the disc - namely tidal shocking (see
e.g. \bcite{1997ApJ...474..223G}, \bcite{1999ApJ...522..935G} and
\bcite{1999ApJ...514..109G}). These tidal shocks can (and in this case
do) become more important than stripping in the very centre of the
satellite. We discuss the effects of tidal shocks in more detail in a
companion paper, \citep{Readinprep1}.

As for the circular orbit case, orbital transformations drive all star
orbits eventually towards the prograde limit at pericentre (blue solid
vertical line). However it is difficult to ascertain whether the
prograde limit is ever actually reached in practice for general
satellite orbits because of the action of tidal shocks. 

\subsection{Implications of the new tidal radii}\label{sec:implications}

A natural question relates to the observability of the three tidal
radii in the projected surface brightness profile
and velocity dispersion of the stars. These are shown in Figure
\ref{fig:realprofiles}. Notice that if only surface
brightness information were available, the tidal radius determined
purely from `features' in the light profile would be erroneous. For
the case of a satellite on a circular orbit (top panels), the
surface brightness profile is depleted at the prograde stripping
radius (blue vertical line) and breaks at the retrograde stripping
radius (red vertical line). For a more
general satellite orbit (bottom panels) there are no observable
features in the light profile at any of the analytic tidal radii. This
is because tidal shocking washes out such information (notice that
such tidal shocks also lower the central surface brightness as the
satellite becomes puffed up). However, if the radial and tangential velocity
dispersions are known then the tidal radii may be much better
determined. Notice that for both the satellite on the circular orbit
and the more general orbit, tangential velocity anisotropy appears at
the prograde stripping radius. This is because, as discussed in
section \ref{sec:tidalbraking}, stars on radial or near radial orbits
are, over long times, pushed momentarily onto prograde orbits, at
which point they are stripped away. Thus even at the prograde stripping
radius, there is a depletion of radial orbits with respect to
retrograde orbits and this leads to the onset of tangential
anisotropy. Such anisotropies increase outwards as stars moving on
radial orbits become more easily stripped at the radial stripping
radius. The presence of tangential
anisotropy provides a much more robust signature of the true tidal
radius than can be measured from the light profile alone.

\section{Conclusions}\label{sec:conclusions}

We have presented an improved analytic calculation for the tidal radius of
satellites and tested our results against N-body simulations.

The tidal radius in general depends upon four factors: 
the potential of the host galaxy, the potential of the satellite, the
orbit of the satellite and {\it the orbit of the star within the
  satellite}. We demonstrated that this last point is 
critical and suggest using {\it three tidal radii} to cover the extrema
range of orbits of stars within the satellite. In this way we
showed explicitly that prograde star orbits will be more easily
stripped than radial orbits; while radial orbits are more easily
stripped than retrograde ones. This result has previously been
established by several authors numerically. We showed further that, in
general, the tidal radius 
{\it does not depend only on the enclosed mean density of the
  satellite}. While for stars on pure radial orbits it is reasonable
to approximate the tidal radius as being that where the density of the
satellite matches that of the host galaxy, for prograde and retrograde
star orbits, the tidal radius also depends on the {\it mass
  distribution} of the satellite galaxy.

Over short times ($\simlt 1-2$\,Gyrs $\sim $1 satellite orbit), we
find excellent agreement between our analytic and numerical models. Over longer
times, star orbits within the satellite are transformed by the tidal
field of the host galaxy. In a Hubble time, this causes a convergence
of the three limiting tidal radii towards the prograde stripping radius.

Satellites which have been tidally
stripped will show tangential velocity anisotropy near their tidal radii, with
a depletion of both prograde and radial orbits relative to the
retrograde orbits. This must be true irrespective of the initial
  conditions. These results naturally 
explain the numerical observations dating back to
\citet{1972ApJ...178..623T} and \citet{1975AJ.....80..290K} which
found exactly these velocity anisotropies in their simulations of
galaxy-satellite interactions.

In the future, with improved kinematic data for nearby satellites from
SIM, it may be possible to search for tangential velocity
anisotropies as a `smoking gun' from tidal stripping, and thereby much
better determine the tidal radii of nearby galaxies and star
clusters. This would represent a significant improvement on current
methods which have historically found `tidal radii' from features in
the surface brightness profiles alone. 

\section{Acknowledgements}
JIR and MIW would like to thank PPARC for grants which have supported
this research. We would like to thank Ben Moore for useful comments
which led to this final version.

\bigskip
\vfil
\bibliographystyle{mn2e}
\bibliography{/home/jir22/More_space__/LaTeX/BibTeX/refs}
 
\end{document}